\documentclass[12pt]{article}

\title{Casper the Friendly Finality Gadget}
\author{
        Vitalik Buterin \textnormal{ and } Virgil Griffith \\        
        Ethereum Foundation}

\usepackage[nonatbib,final]{nips_2017}

\usepackage{color}
\usepackage{graphicx}
\DeclareGraphicsExtensions{.pdf,.eps,.png,.jpg}		% search for .pdf, then .eps, then .pngs, then .jpg

\graphicspath{{figs/}{figures/}{images/}{./}}

\usepackage{url}
\usepackage{amsmath}

\usepackage{subcaption}

\usepackage{array}			% replacement for eqnarray.  Must be BEFORE \usepackage{arydshln}
\usepackage{units}			% for \nicefrac{\alpha}{\beta}

\usepackage{amsthm}		% for theorems

\usepackage{microtype}

\usepackage{wasysym}

\usepackage{textcomp, marvosym} % pretty symbols
\usepackage{booktabs} 	% for much better looking tables

\usepackage{dsfont}

\usepackage{titlesec,titlecaps}

\Addlcwords{is with of the and in}
\Addlcwords{of the}
\Addlcwords{and}
\titleformat{\section}[block]{}{\normalfont\Large\bfseries\thesection.\;}{0pt}{\formatsectiontitle}
\newcommand{\formatsectiontitle}[1]{\normalfont\Large\bfseries\titlecap{#1}}

\titleformat{\subsection}[block]{}{\normalfont\large\bfseries\thesubsection.\;}{0pt}{\formatsubsectiontitle}
\newcommand{\formatsubsectiontitle}[1]{\normalfont\large\bfseries\titlecap{#1}}

\usepackage{float} % for \subfloat

\usepackage{xspace}

\renewcommand{\eqref}[1]{eq.~(\ref{#1})}

\newcommand{\figref}[1]{Figure~\ref{#1}}

\newtheorem{theorem}{Theorem}

\usepackage{makecell}

\renewcommand*{\to}{\rightarrow}

\usepackage{mdframed}

\usepackage{hyperref}
\usepackage{upgreek}
\usepackage{datetime}

\newcommand{\signature}{\ensuremath{\mathcal{S}}\xspace}

\newcommand{\h}{\operatorname{h}\xspace}

\newcommand{\DS}{\operatorname{DS}}
\newcommand{\DE}{\operatorname{DE}}

\newcommand{\msgVOTE}{\textbf{\textsc{vote}}\xspace}

\begin{document}

\maketitle
%\TODO{\vspace{-0.2in} \today\ \ \currenttime}

%%%%%%%%%%%%%%%%%%%%%%%%%%%%%%%%%%%%%%%%%%%%%%%%%%%%%%%%%%%%%%%%%%%%%%%%%%%%%%%%%%%%%%%%%%%%%%%%%%%%%%%%%%%%%%%%%%%%%%%%%%
%%%%%%%%%%%%%%%%%%%%%%%%%%%%%%%%%%%%%%%%%%%%%%%%%%%%%%%%%%%%%%%%%%%%%%%%%%%%%%%%%%%%%%%%%%%%%%%%%%%%%%%%%%%%%%%%%%%%%%%%%%

\begin{abstract}
We introduce Casper, a proof of stake-based finality system which overlays an existing proof of work blockchain. Casper is a partial consensus mechanism combining proof of stake algorithm research and Byzantine fault tolerant consensus theory.  We introduce our system, prove some desirable features, and show defenses against long range revisions and catastrophic crashes.  The Casper overlay provides almost any proof of work chain with additional protections against block reversions.
\end{abstract}

%%%%%%%%%%%%%%%%%%%%%%%%%%%%%%%%%%%%%%%%%%%%%%%%%%%%%%%%%%%%%%%%%%%%%%%%%%%%%%%%%%%%%%%%%%%%%%%%%%%%%%%%%%%%%%%%%%%%%%%%%%
%%%%%%%%%%%%%%%%%%%%%%%%%%%%%%%%%%%%%%%%%%%%%%%%%%%%%%%%%%%%%%%%%%%%%%%%%%%%%%%%%%%%%%%%%%%%%%%%%%%%%%%%%%%%%%%%%%%%%%%%%%
\section{Introduction}
\label{sect:intro}

Over the past few years there has been considerable research into ``proof of stake'' (PoS) based blockchain consensus algorithms. In a PoS system, a blockchain appends and agrees on new blocks through a process where anyone who holds coins inside of the system can participate, and the influence an agent has is proportional to the number of coins (or ``stake'') it holds. This is a vastly more efficient alternative to proof of work (PoW) ``mining'' and enables blockchains to operate without mining's high hardware and electricity costs.

There are two major schools of thought in PoS design. The first, \textit{chain-based proof of stake}\cite{shi2017,dfinity}, mimics proof of work mechanics and features a chain of blocks and simulates mining by pseudorandomly assigning the right to create new blocks to stakeholders.  This includes Peercoin\cite{king2012ppcoin}, Blackcoin\cite{vasin2014blackcoin}, and Iddo Bentov's work\cite{bentov2016pos}.

The other school, \textit{Byzantine fault tolerant} (BFT) based proof of stake, is based on a thirty-year-old body of research into BFT consensus algorithms such as PBFT\cite{castro1999practical}. BFT algorithms typically have proven mathematical properties; for example, one can usually mathematically prove that as long as $>\!\frac{2}{3}$ of protocol participants are following the protocol honestly, then, regardless of network latency, the algorithm cannot finalize conflicting blocks.  Repurposing BFT algorithms for proof of stake was first introduced by Tendermint\cite{kwon2014tendermint}, and has modern inspirations such as \cite{algorand}.  Casper follows this BFT tradition, though with some modifications.

\subsection{Our Work}
Casper the Friendly Finality Gadget is an overlay atop a \emph{proposal mechanism}---a mechanism which proposes blocks\footnote{This functionality serves a similar role to the common abstraction of ``leader election'' used in traditional BFT algorithms, but is adapted to accomodate Casper's construction of being a finality overlay atop an existing blockchain.}.  Casper is responsible for finalizing these blocks, essentially selecting a
unique chain which represents the canonical transactions of the ledger.  Casper provides safety, but liveness depends on the chosen proposal mechanism.  That is, if attackers wholly control the proposal mechanism, Casper protects against finalizing two conflicting checkpoints, but the attackers could prevent Casper from finalizing any future checkpoints.

Casper introduces several new features that BFT algorithms do not necessarily support:
\begin{itemize}
\item \textbf{Accountability}.  If a validator violates a rule, we can detect the violation and know which validator violated the rule.  Accountability allows us to penalize malfeasant validators, solving the ``nothing at stake'' problem that plagues chain-based PoS. The penalty for violating a rule is a validator's entire deposit.  This maximal penalty is the defense against violating the protocol.  Because proof of stake security is based on the size of the penalty, which can be set to greatly exceed the gains from the mining reward, proof of stake provides strictly stronger security incentives than proof of work.

\item \textbf{Dynamic validators}. We introduce a safe way for the validator set to change over time (Section \ref{sect:join_and_leave}).

\item \textbf{Defenses}.  We introduce defenses against long range revision attacks as well as attacks where more than $\frac{1}{3}$ of validators drop offline, at the cost of a very weak tradeoff synchronicity assumption (Section \ref{sect:attacks}).

\item \textbf{Modular overlay}.  Casper's design as an overlay makes it easier to implement as an upgrade to an existing proof of work chain.
\end{itemize}

We describe Casper in stages, starting with a simple version (Section \ref{sect:protocol}) and then progressively adding validator set changes (Section \ref{sect:join_and_leave}) and finally defenses against attacks (Section \ref{sect:attacks}).

%%%%%%%%%%%%%%%%%%%%%%%%%%%%%%%%%%%%%%%%%%%%%%%%%%%%%%%%%%%%%%%%%%%%%%%%%%%%%%%%%%%%%%%%%%%%%%%%%%%%%%%%%%%%%%%%%%%%%%%%%%
%%%%%%%%%%%%%%%%%%%%%%%%%%%%%%%%%%%%%%%%%%%%%%%%%%%%%%%%%%%%%%%%%%%%%%%%%%%%%%%%%%%%%%%%%%%%%%%%%%%%%%%%%%%%%%%%%%%%%%%%%%

\section{The Casper Protocol}
\label{sect:protocol}

Within Ethereum, the proposal mechanism will initially be the existing proof of work chain, making the first version of Casper a hybrid PoW/PoS system.  In future versions the PoW proposal mechanism will be replaced with something more efficient.  For example, we can imagine converting the block proposal into a some kind of PoS round-robin block signing scheme.

In this simple version of Casper, we assume there is a fixed set of validators and a proposal mechanism (e.g., the familiar proof of work proposal mechanism) which produces child blocks of existing blocks, forming an ever-growing \emph{block tree}.  From \cite{nakamoto} the root of the tree is typically called the ``genesis block''.

%Note that for an individual block, there will often be multiple blocks added as new children by this mechanism, so the block tree will have many branches.

Under normal circumstances, we expect that the proposal mechanism will typically propose blocks one after the other in a linked list (i.e., each ``parent'' block having exactly one ``child'' block).  But in the case of network latency or deliberate attacks, the proposal mechanism will inevitably occassionally produce multiple children of the same parent. Casper's job is to choose a single child from each parent, thus choosing one canonical chain from the block tree.

\begin{figure}[htb]
\centering
   \begin{subfigure}[b]{0.90\textwidth}
   \centering
   \includegraphics[width=2.8in]{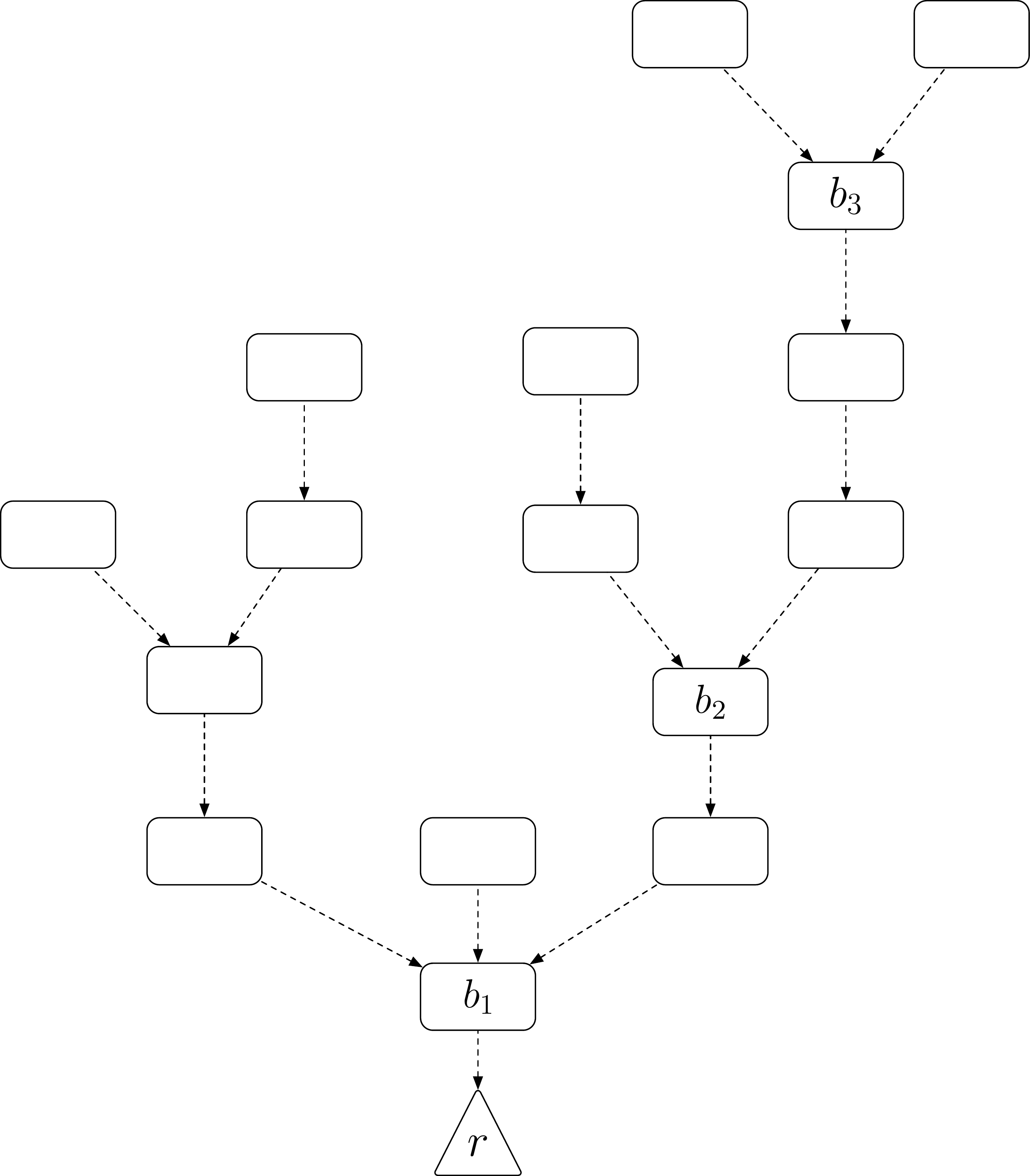}
	\caption{The checkpoint tree.  The dashed line represents 99 blocks between the checkpoints, which are represented by rounded rectangles.  The root of the tree is denoted ``r''.}
	\label{fig:2a}	
	\end{subfigure}
	\vspace{0.2in}
	
\begin{subfigure}[b]{0.45\textwidth}
   \centering
   \includegraphics[height=2.8in]{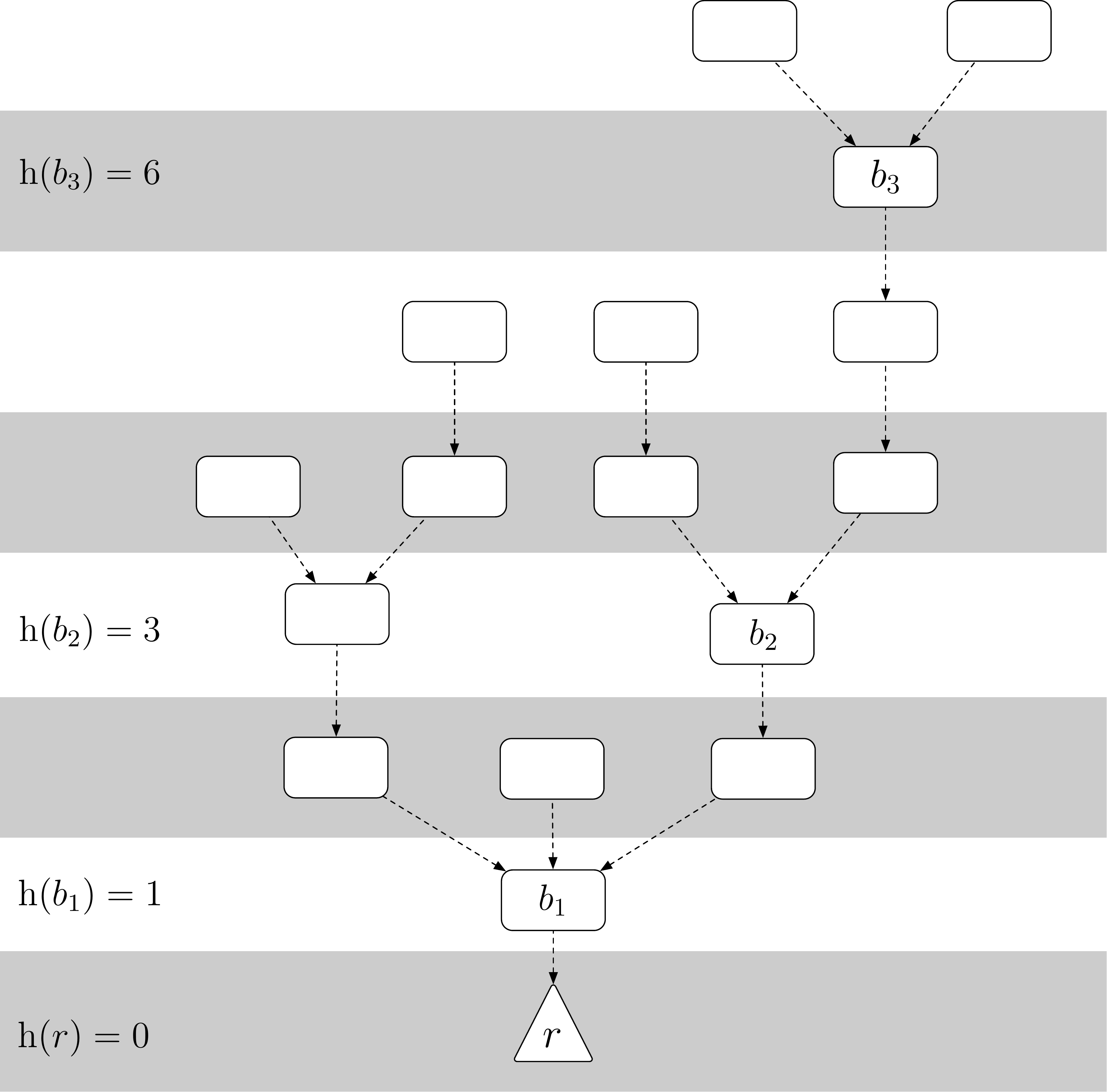}
	\caption{The height function}
	\label{fig:2b}	
	\end{subfigure} \hspace{0.05\textwidth} 	 \begin{subfigure}[b]{0.45\textwidth}
   \centering
   \includegraphics[height=2.8in]{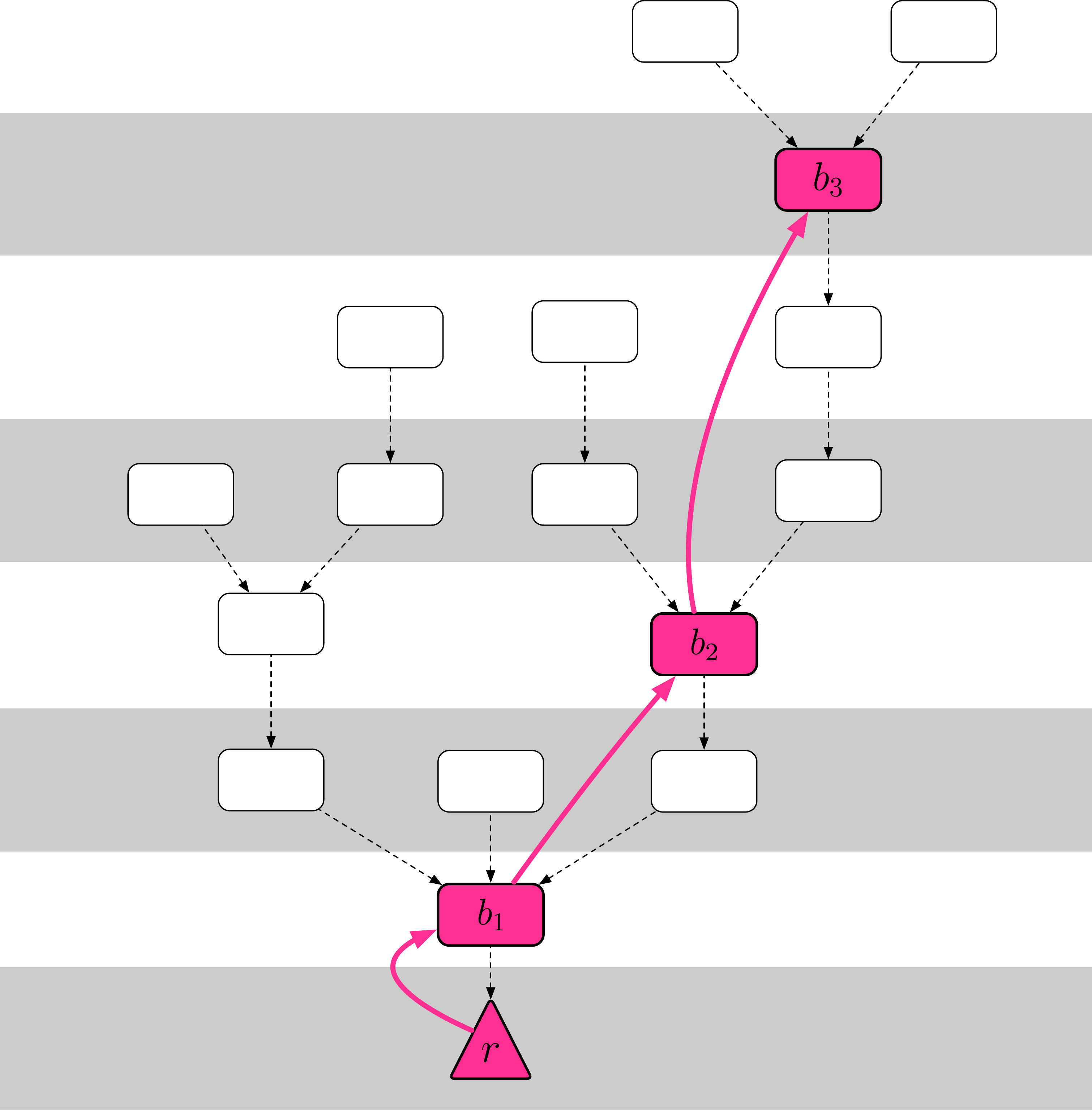}
%	\caption{The justified chain $(r, b_1, b_2, b_3)$}
	\caption{The justified chain $r \to b_1 \to b_2 \to b_3$}
	\label{fig:2c}	
	\end{subfigure}

\caption{Illustrating a checkpoint tree, the height function, and a justified chain within the checkpoint tree.}
\label{fig:conflicting_checkpoints}
\end{figure}

Rather than deal with the full block tree, for efficiency purposes\footnote{A long distance between checkpoints reduces the overhead of the algorithm, but also increases the time it takes to come to consensus.  We choose a spacing of 99 blocks between checkpoints as a middle ground.} Casper only considers the subtree of \emph{checkpoints} forming the \emph{checkpoint tree} (\figref{fig:2a}).  The genesis block is a checkpoint, and every block whose height in the block tree (or “block number”) is an exact multiple of 100 is also a checkpoint. The ``checkpoint height'' of a block with block height $100 * k$ is simply $k$; equivalently, the height $\h(c)$ of a checkpoint $c$ is the number of elements in the checkpoint chain stretching from $c$ (non-inclusive) to the root along the parent links (\figref{fig:2b}).\footnote{Specifically, the height of a checkpoint is \emph{not} the same as the number of ancestors in the checkpoint tree all the way back to root along the supermajority links (defined in the next section).}

Each validator has a \emph{deposit}; when a validator joins, its deposit is the number of deposited coins.  After joining, each validator's deposit rises and falls with rewards and penalties.  Proof of stake's security derives from the size of the deposits, not the number of validators, so for the rest of this paper, when we say ``$\frac{2}{3}$ of validators'', we are referring to the \emph{deposit-weighted} fraction; that is, a set of validators whose sum deposit size equals to $\frac{2}{3}$ of the total deposit size of the entire set of validators.

Validators can broadcast a \textit{vote} message containing four pieces of information (Table \ref{tbl:messages}): two checkpoints $s$ and $t$ together with their heights $\h(s)$ and $\h(t)$.  We require that $s$ be an ancestor of $t$ in the checkpoint tree, otherwise the vote is considered invalid.  If the public key of the validator $\upnu$ is not in the validator set, the vote is considered invalid.  Together with the signature of the validator, we will write these votes in the form $\left\langle \upnu, s, t, \h(s), \h(t) \right\rangle$.

\begin{table}[bth]
\centering

   \begin{tabular}{l l}
	\toprule
	\textbf{Notation} & \textbf{Description} \\
	\midrule
	$s$ & the hash of any justified checkpoint (the ``source'') \\
	$t$ & any checkpoint hash that is a descendent of  $s$ (the ``target'') \\
	$\h(s)$ & the height of checkpoint $s$ in the checkpoint tree \\
	$\h(t)$ & the height of checkpoint $t$ in the checkpoint tree \\
	\signature & signature of $\left\langle s, t, \h(s), \h(t) \right\rangle$ from the validator $\upnu$'s private key \\
	\bottomrule
	\end{tabular}

\vspace{0.15in}
\caption{The schematic of a single \msgVOTE message denoted $\left\langle \upnu, s, t, \h(s), \h(t) \right\rangle$.}
\label{tbl:messages}
\end{table}

We define the following terms:
\begin{itemize}
\item A \emph{supermajority link} is an ordered pair of checkpoints $(a, b)$, also written $a \rightarrow b$, such that at least $\frac{2}{3}$ of validators (by deposit) have published votes with source $a$ and target $b$.  Supermajority links \emph{can skip checkpoints}, i.e., it's perfectly okay for $\h(b) > \h(a) + 1$.  \figref{fig:2c} shows three distinct supermajority links in red: $r \to b_1$, $b_1 \to b_2$, and $b_2 \to b_3$.

\item Two checkpoints $a$ and $b$ are called \emph{conflicting} if and only if they are nodes in distinct branches, i.e., neither is an ancestor or descendant of the other.

\item A checkpoint $c$ is called \emph{justified} if (1) it is the root, or (2) there exists a supermajority link $c^\prime \to c$ where checkpoint $c^\prime$ is justified.  \figref{fig:2c} shows a chain of four justified blocks.

\item A checkpoint $c$ is called \emph{finalized} if (1) it is the root or (2) it is justified and there is a supermajority link $c \to c^\prime$ where $c^\prime$ is a \emph{direct child} of $c$.  Equivalently, checkpoint $c$ is finalized if and only if: checkpoint $c$ is justified, there exists a supermajority link $c \to c^\prime$, checkpoints $c$ and $c^\prime$ are not conflicting, and $\h(c^\prime) = \h(c) + 1$.
\end{itemize}

The most notable property of Casper is that it is impossible for any two  conflicting checkpoints to be finalized unless $\geq \frac{1}{3}$ of the validators violate one of the two\footnote{Earlier versions of Casper had two types of messages and four slashing conditions\cite{minslashing}, but we have reduced this to one message type and two slashing conditions.  We removed the conditions: (i) Committed hashes must already be justified, and (ii) prepare messages must point to an already justified ancestor.  This is a design choice.  We made this choice so that slashing violations are independent of the state of the chain.} Casper Commandments/slashing conditions (\figref{fig:commandments}).

%\clearpage

\begin{figure}
\begin{mdframed}
\textsc{An individual validator $\upnu$ must not publish two distinct votes,}
\begin{equation*}
\left\langle \upnu, s_1, t_1, \h(s_1), \h(t_1)\right\rangle \hspace{0.5in} \textsc{and} \hspace{0.5in} \left\langle \upnu, s_2, t_2, \h(s_2), \h(t_2)\right\rangle \; ,
\end{equation*}
\textsc{such that either:}

\begin{enumerate}
   \item[\textbf{I.}] $\h(t_1) = \h(t_2)$.

   Equivalently, a validator must not publish two distinct votes for the same target height.
\end{enumerate}
\vspace{-0.15in}
\textsc{or}

\begin{enumerate}
   \item[\textbf{II.}] $\h(s_1) < \h(s_2) < \h(t_2) < \h(t_1)$.

   Equivalently, a validator must not vote within the span of its other votes.
\end{enumerate}
\end{mdframed}
\caption{The two Casper Commandments.  Any validator who violates either of these commandments gets its deposit slashed.}
\label{fig:commandments}
\end{figure}

If a validator violates either slashing condition, the evidence of the violation can be included into the blockchain as a transaction, at which point the validator's entire deposit is taken away with a small ``finder's fee'' given to the submitter of the evidence transaction. In current Ethereum, stopping the enforcement of a slashing condition requires a successful $51\%$ attack on Ethereum's proof-of-work block proposer.

%%%%%%%%%%%%%%%%%%%%%%%%%%%%%%%%%%%%%%%%%%%%%%%%%%%%%%%%%%%%%%%%%%%%%%%%%%%%%%%%%%%%%%%%%%%%%%%%%%%%%%%%%%%%%%%%%%%%%%%%%%
%%%%%%%%%%%%%%%%%%%%%%%%%%%%%%%%%%%%%%%%%%%%%%%%%%%%%%%%%%%%%%%%%%%%%%%%%%%%%%%%%%%%%%%%%%%%%%%%%%%%%%%%%%%%%%%%%%%%%%%%%%
\subsection{Proving Safety and Plausible Liveness}
\label{sect:theorems}

We prove Casper's two fundamental properties: \textit{accountable safety} and \textit{plausible liveness}. Accountable safety means that two conflicting checkpoints cannot both be finalized unless $\geq \frac{1}{3}$ of validators violate a slashing condition (meaning at least one third of the total deposit is lost).  Plausible liveness means that, regardless of any previous events (e.g., slashing events, delayed blocks, censorship attacks, etc.), if $\geq \frac{2}{3}$ of validators follow the protocol, then it's always possible to finalize a new checkpoint without any validator violating a slashing condition.

\emph{Under the assumption that $< \frac{1}{3}$ of the validators by weight violate a slashing condition}, we have the following properties:

\begin{enumerate}
    \item[(i)] If $s_1 \rightarrow t_1$ and $s_2 \rightarrow t_2$ are distinct supermajority links, then $\h(t_1) \not= \h(t_2)$.
    \item[(ii)] If $s_1 \rightarrow t_1$ and $s_2 \rightarrow t_2$ are distinct supermajority links, then the inequality $\h(s_1)~<~\h(s_2)~<~\h(t_2)~<~\h(t_1)$ cannot hold.
\end{enumerate}

From these two properties, we can immediately see that, for any height $n$: 
\begin{enumerate}
\item[(iii)] there exists at most one supermajority link $s \to t$ with $\h(t) = n$.
\item[(iv)] there exists at most one justified checkpoint with height $n$.  
\end{enumerate}

With these four properties in hand, we move to the main theorems.

\begin{figure}[h!tb]
\centering
   \includegraphics[width=4in]{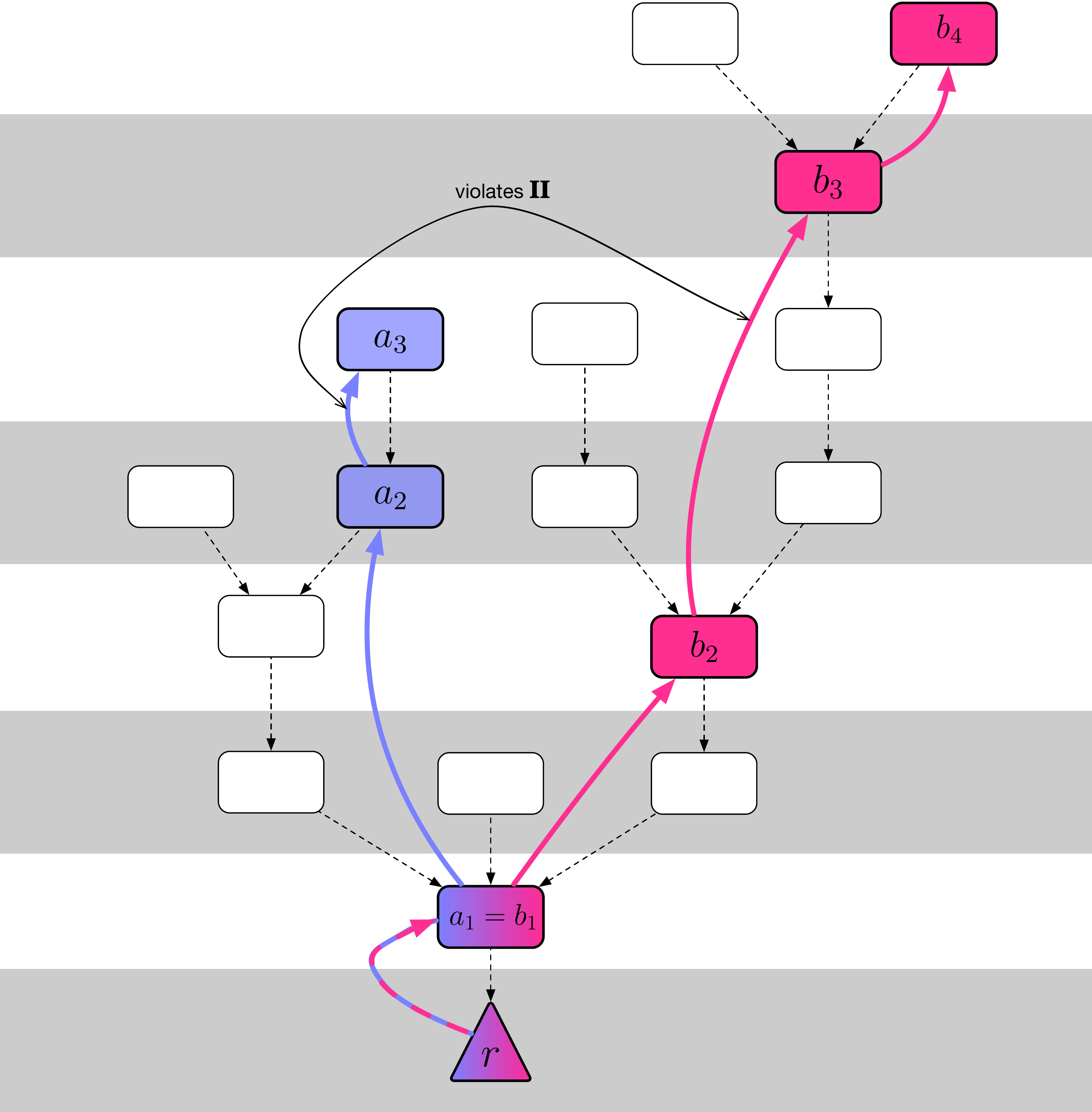}
\caption{Figure for Theorem \ref{theorem:safety} (Accountable Safety).}
\label{fig:proof1}	
\end{figure}

\begin{theorem}[Accountable Safety]
\label{theorem:safety}
Two conflicting checkpoints $a_m$ and $b_n$ cannot both be finalized.
\begin{proof}
Let $a_m$ (with justified direct child $a_{m+1}$, meaning $\h(a_m) + 1 = \h(a_{m+1})$) and $b_n$ (with justified direct child $b_{n+1}$, meaning $\h(b_n) + 1 = \h(b_{n+1})$) be distinct finalized checkpoints as in \figref{fig:proof1}. Now suppose $a_m$ and $b_n$ conflict, and without loss of generality $\h(a_m) < \h(b_n)$ (if $\h(a_m) = \h(b_n)$, then it is clear that $\frac{1}{3}$ of validators violated condition \textbf{I}). Let $r \rightarrow b_1 \rightarrow b_2 \rightarrow \cdots \rightarrow b_n$ be a chain of checkpoints, such that there exists a supermajority link $r \to b_1$, $\ldots$, $b_i \to b_{i+1}, \ldots, b_{n} \to b_{n+1}$.  We know that no $\h(b_i)$ equals either $\h(a_m)$ or $\h(a_{m+1})$, because that violates property (iv). Let $j$ be the lowest integer such that $\h(b_j) > \h(a_{m+1})$; then $\h(b_{j-1}) < \h(a_{m+1})$ (or $\h(b_{j-1}) = \h(a_m)$, which violates condition \textbf{I}). This implies the existence of a supermajority link from a checkpoint with an epoch number less than $\h(a_m)$ to a checkpoint with an epoch number greater than $\h(a_{m+1})$, which is incompatible with the supermajority link from $a_m$ to $a_{m+1}$.
\end{proof}
\end{theorem}
% $(r, b_1, b_2, \ldots, b_n)$ 

\begin{theorem}[Plausible Liveness]
\label{theorem:liveness}
Supermajority links can always be added to produce new finalized checkpoints, provided there exist children extending the finalized chain.\footnote{The finalized chain always exists because the root is  finalized by definition.}
\begin{proof}
Let $a$ be the justified checkpoint with the greatest height, and $b$ be the target checkpoint with the greatest height for which any validator has made a vote. Any checkpoint $a^\prime$ which is a descendant of $a$ with height $\h(a^\prime) = \h(b) + 1$ can be justified without violating either commandments \textbf{I} or \textbf{II}, and then $a^\prime$ can be finalized by adding a supermajority link from $a^\prime$ to a direct child of $a^\prime$, also without violating either \textbf{I} or \textbf{II}.
\end{proof}

\end{theorem}

\subsection{Casper's Fork Choice Rule}
\label{sect:forkchoice}

Casper is more complicated than standard PoW designs.  As such, the fork-choice must be adjusted.  Our modified fork-choice rule should be followed by all users, validators, and even the underlying block proposal mechanism.  If the users, validators, or block-proposers instead follow the standard PoW fork-choice rule of ``always build atop the longest chain'', there are pathological scenarios where Casper gets ``stuck'' and any blocks built atop the longest chain cannot be finalized (or even justified) without some validators altruistically sacrificing their deposit.  To avoid this, we introduce a novel, correct by construction, fork choice rule: \textsc{Follow the chain containing the justified checkpoint of the greatest height}.  This fork choice rule is correct by construction because it follows from the plausible liveness proof (Theorem \ref{theorem:liveness}), which precisely states that it's always possible to finalize a new checkpoint on top of the justified checkpoint with the greatest height.  This fork choice rule will be tweaked in Sections \ref{sect:join_and_leave} and \ref{sect:attacks}.

\section{Enabling Dynamic Validator Sets}
\label{sect:join_and_leave}

The set of validators needs to be able to change.  New validators must be able to join, and existing validators must be able to leave.  To accomplish this, we define the \textit{dynasty} of a block.  The \emph{dynasty of block $b$} is the number of finalized checkpoints in the chain from root to the parent of block $b$.  When a would-be validator's \emph{deposit message} is included in a block with dynasty $d$, then the validator $\upnu$ will join the validator set at first block with dynasty $d+2$.  We call $d+2$ this validator's \textit{start dynasty}, $\DS(\upnu)$.

To leave the validator set, a validator must send a ``withdraw'' message. If validator $\upnu$'s  withdraw message is included in a block with dynasty $d$, it similarly leaves the validator set at the first block with dynasty $d+2$; we call $d+2$ the validator's \textit{end dynasty}, $\DE(\upnu)$.  If a withdraw message has not yet been included, then $\DE(\upnu) = \infty$.  Once validator $\upnu$ leaves the validator set, the validator's public key is forever forbidden from rejoining the validator set.  This removes the need to handle multiple start/end dynasties for a single identifier.

At the start of the end dynasty, the validator's deposit is locked for a long period of time, called the \textit{withdrawal delay} (approximately ``four months worth of blocks''), before the deposit is withdrawn.  If, during the withdrawal delay, the validator violates any commandment, the deposit is slashed.

%We define $\mathbf{V}$ as the set of all validators who have sever joined the validator set. Then, 
We define two functions that generate two subsets of validators for any given dynasty $d$, the \emph{forward validator set} and the \emph{rear validator set}.  They are defined as,

\begin{align*}
    \mathcal{V}_{\operatorname{f}}(d) &\equiv \left\{ \upnu : \DS(\upnu) \leq d < \DE(\upnu)  \right\} \\
    \mathcal{V}_{\operatorname{r}}(d) &\equiv \left\{ \upnu : \DS(\upnu) < d \leq \DE(\upnu) \right\} \; .
\end{align*}

Note this means that the forward validator set of dynasty $d$ is the rear validator set of dynasty $d+1$, i.e., $\mathcal{V}_{\operatorname{f}}(d) = \mathcal{V}_{\operatorname{r}}(d+1)$.  To support these dynamic validator sets, we redefine a supermajority link and a finalized checkpoint as follows:
\begin{itemize}

\item An ordered pair of checkpoints $(s, t)$, where $t$ is in dynasty $d$, has a \emph{supermajority link} if both at least $\frac{2}{3}$ of the forward validator set of dynasty $d$ have published votes $s \rightarrow t$ and at least $\frac{2}{3}$ of the rear validator set of dynasty $d$ have published votes $s \rightarrow t$.

\item Previously, a checkpoint $c$ was called \emph{finalized} if $c$ is justified and there is a supermajority link from $c \to c^\prime$ where $c^\prime$ is a child of $c$.  We now add the condition that $c$ is finalized if only if the votes for the supermajority link $c \rightarrow c^\prime$, as well as the supermajority link justifying $c$, are included in $c^\prime$'s block chain and before the child of $c^\prime$---i.e., before block number $\h(c^\prime) * 100 + 1$.
\end{itemize}

%Note that in order for the chain to be able to ``know'' its own current dynasty, we need to restrict our definition of ``finalization'' slightly:		

%Before, a checkpoint $c$ is called \emph{finalized} if it is justified and there is a supermajority link from $c$ to any of its direct children in the checkpoint tree.  Now, finalization has one additional condition---$c$ is finalized only if the votes for the supermajority link $c \rightarrow c^\prime$, as well as all of the supermajority links \hyperlink{recursivelyjustify}{recursively justifying} $c$, are included in the block chain before the child of $c^\prime$, i.e., before block number $\h(c^\prime) * 100$.		

\begin{figure}[htb]
\centering
\includegraphics[height=3in]{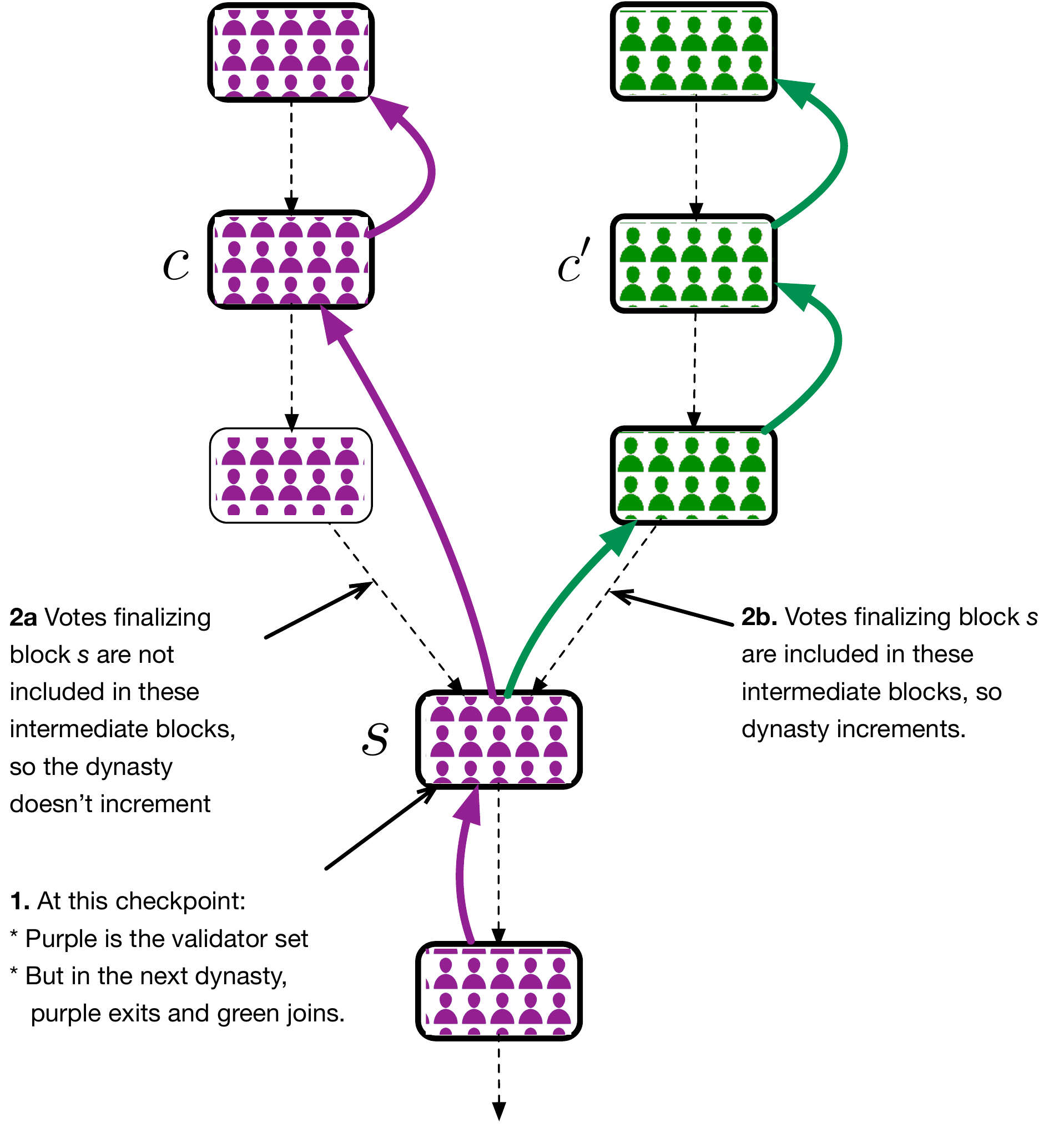}
\caption{\textbf{Attack from dynamic validator sets}.  Without the validator set stitching mechanism, it's possible for two conflicting checkpoints $c$ and $c^\prime$ to both be finalized without any validator getting  slashed.  In this case $c$ and $c^\prime$ are the same height thus violating commandment \textbf{I}, but because the validator sets finalizing $c$ and $c^\prime$ are disjoint, no one gets slashed.}
\label{fig:DYN}
\end{figure}

The forward and rear validator sets will usually greatly overlap; but if the two validator sets substantially differ, this ``stitching'' mechanism prevents safety failure in the case when two grandchildren of a finalized checkpoint have different dynasties because the evidence was included in one chain but not the other.  For an example of this, see \figref{fig:DYN}.

\section{Stopping attacks}
\label{sect:attacks}
There are two well-known attacks against proof-of-stake systems: \emph{long range revisions} and \emph{catastrophic crashes}.  We discuss each in turn.

\subsection{Long Range Revisions}
\begin{figure}[tb]
\centering
\includegraphics[width=3in]{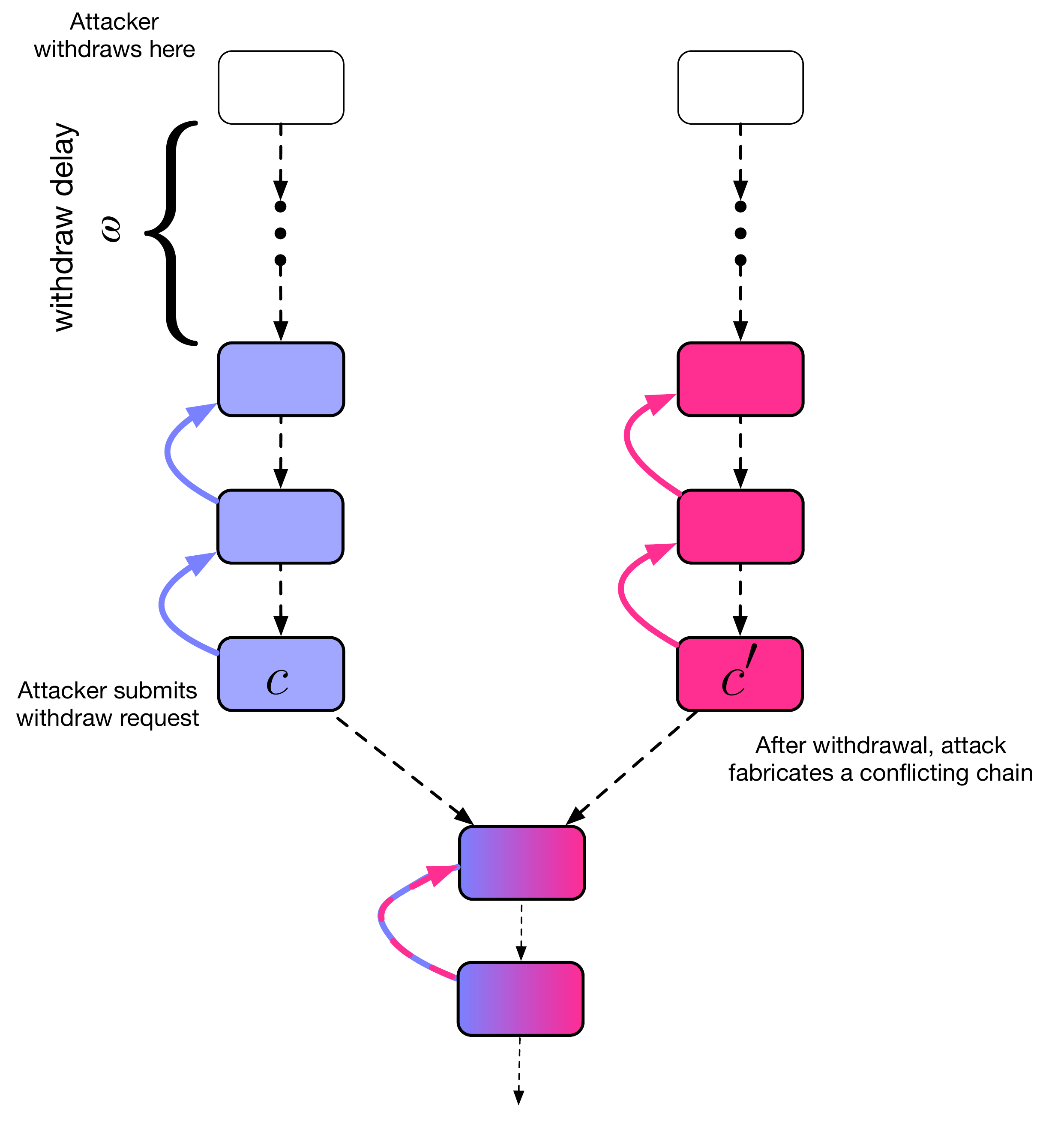}
\caption{\textbf{The long range attack}.  As long as a client gains complete knowledge of the justified chain at a regular interval, it will not be susceptible to a long range attack.}
\label{fig:longrange}
\end{figure}

The withdrawal delay after a validator's end dynasty introduces a synchronicity assumption between validators and clients. Once a coalition of validators has withdrawn their deposits, if that coalition had more than $\frac{2}{3}$ of deposits \emph{long ago in the past}, they can use their historical supermajority to finalize conflicting checkpoints without fear of getting slashed (because they have already withdrawn their money).  This is called the \emph{long-range revision attack}, see in \figref{fig:longrange}.

In simple terms, long-range attacks are prevented by a fork choice rule to never revert a finalized block, as well as an expectation that each client will ``log on'' and gain a complete up-to-date view of the chain at some regular frequency (e.g., once per 1--2 months). A ``long range revision'' fork that finalizes blocks older than that will simply be ignored, because all clients will have already seen a finalized block at that height and will refuse to revert it.

We make an informal proof of the mechanism as follows. Suppose that:

\begin{itemize}

\item There is a maximum communication delay $\delta$ between two clients, so if one client hears some message at time $t$, all other clients are guaranteed to have heard it by time $t + \delta$. This means that we can talk about the ``time window'' $[t_{\min}, t_{\max}]$ during which a block was received by the network, with width $t_{\max} - t_{\min}$ at most $\delta$.
\item We assume that all clients have local clocks that are perfectly synchronized (any discrepancy can be treated as being part of the communication delay $\delta$).
\item Blocks are required to have timestamps. If a client has local time $T_L$, then it will reject blocks whose timestamp $T_B$ satisfies $T_B > T_L$ (i.e., in the future), and they will refuse to accept as finalized (but may still accept as part of the chain) blocks where $T_B < T_L - \delta$ (i.e., too far in the past)
\item If a validator sees a slashing violation at time $t$ (that's the time they hear the \emph{later} of the two votes required for a slashing violation), then they reject blocks with timestamps $> t + 2\delta$ that are part of chains that have not yet included this slashing evidence.
\end{itemize}

Suppose that a large set of slashing violations results in two conflicting finalized checkpoints, $c_1$ and $c_2$.  If the two time windows do not intersect, then all validators agree which checkpoint came first and everyone follows the rule to not revert finalized checkpoints then there is no issue.

If the two time windows \emph{do} intersect, then we can handle the case as follows. Let $c_1$'s time window be $[0, \delta]$ and $c_2$'s  time window be $[\delta - \epsilon, 2\delta - \epsilon]$. Then the timestamps of both are at least $0$. By time $2\delta$ it is guaranteed that all clients have seen the slashing violation, so they reject blocks with timestamp $> 4\delta$ whose chains have not yet included the evidence transaction. Hence, as long as $\omega > 4\delta$, it's guaranteed that malfeasant validators will lose their deposits in all chains that any client accepts, where $\omega$ is the ``withdraw delay'' (\figref{fig:longrange}), the delay between the end-epoch and when validators actually receive their deposits back.

Due to network delays, it's possible that clients will disagree whether a given piece of slashing evidence was submitted into a given chain ``on time'' or as having accepted it too late.  However, this is only a liveness failure, not a safety failure, and this possibility does not weaken our security claims because it is already known that a corrupted proposal mechanism (which would be required to prevent evidence inclusion) can prevent finality.

We can also sidestep the issue of evidence inclusion timeouts by informally arguing that attacks  will be short-lived, because the validators will perceive a long-running chain without including slashing evidence as an attack and switch to another branch supported by an honest minority of validators that are not part of the attack (see Section \ref{sect:leak}) thus stopping the attack and slashing the attacker.

\subsection{Castastrophic Crashes}
\label{sect:leak}

Suppose that $>\frac{1}{3}$ of validators crash-fail at the same time---i.e., they are no longer connected to the network due to a network partition, computer failure, or the validators themselves are malicious. Intuitively, from this point on, no supermajority links can be created, and thus no future checkpoints can be finalized.

We can recover from this by instituting an ``inactivity leak'' which slowly drains the deposit of any validator that does not vote for checkpoints, until eventually its deposit sizes decrease low enough that the validators who \emph{are} voting are a supermajority.  The simplest formula is something like ``in every epoch a validator with deposit size $D$ fails to vote, it loses $D * p$ (for $0 < p < 1$)'', though to resolve catastrophic crashes more quickly a formula which increases the leak rate in the event of a long streak of non-finalized blocks may be optimal.

This drained ether can be burned or returned to the validator after $\omega$ days.  Whether leaked assets should be burned or returned as well as  the exact formula for the inactivity leak is outside the scope of this paper as these are questions of economic incentives, not Byzantine-fault-tolerance.

\begin{figure}[htb]
\centering
\includegraphics[width=3in]{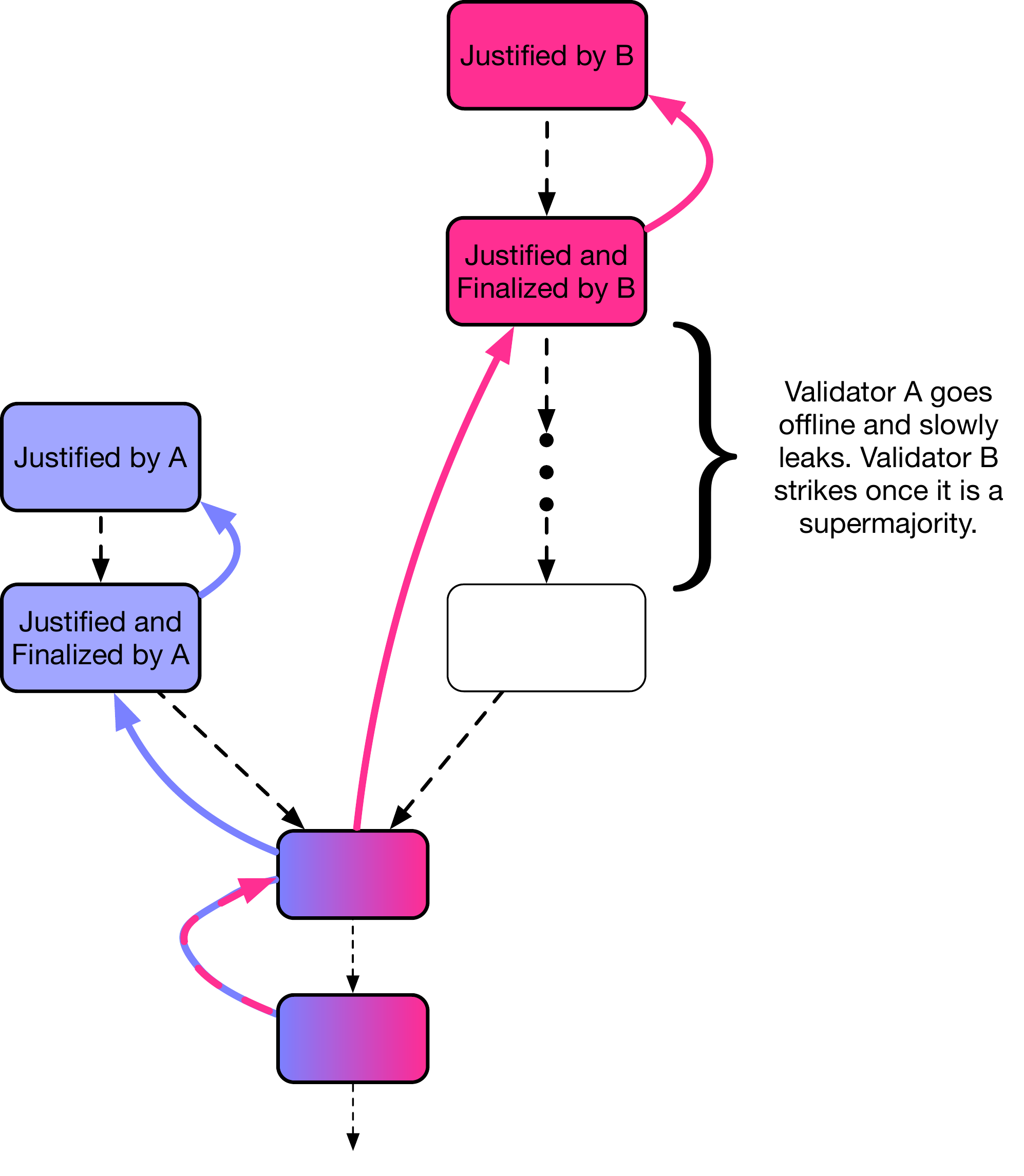}
\caption{\textbf{Inactivity leak}.  The checkpoint on the left can be finalized immediately. The checkpoint on the right can be finalized after some time, once offline validator deposits sufficiently deplete.}
\label{fig:bigcrash}
\end{figure}

The inactivity leak introduces the possibility of two conflicting checkpoints being finalized without any validator getting slashed (as in Figure \ref{fig:bigcrash}), with validators only losing money on only one of the two checkpoints. Assume the validators are split into two subsets, with subset $\mathcal{V}_A$ voting on chain $A$ and subset $\mathcal{V}_B$ voting on chain $B$.  On chain $A$, $\mathcal{V}_B$'s deposits will leak, and vice versa, leading to each subset having a supermajority on its respective chain, allowing two conflicting checkpoints to be finalized without any validators being explicitly slashed (but each subset will lose a large portion of their deposit on one of the two chains due to leaks).  If this situation happens, then each validator should simply favor whatever finalized checkpoint it saw first.

The exact algorithm for recovering from these various attacks remains an open problem.  For now, we assume validators can detect obviously malfeasant behavior (e.g., not including evidence) and manually create a ``minority soft fork''.  This minority fork can be viewed as a blockchain in its own right that competes with the majority chain in the market, and if the majority chain truly is operated by colluding malicious attackers then we can assume that the market will favor the minority fork.

\section{Conclusions}
We presented Casper, a novel proof of stake system derived from the Byzantine fault tolerance literature.  Casper includes: two slashing conditions, a correct-by-construction fork choice rule inspired by \cite{sompolinsky2013accelerating}, and dynamic validator sets.  Finally we introduced extensions to Casper (not reverting finalized checkpoints and the inactivity leak) to defend against two common attacks.

Casper remains imperfect.  For example, a wholly compromised block proposal mechanism will prevent Casper from finalizing new blocks.  Casper is a PoS-based strict security improvement to almost any PoW chain.  The problems that Casper does not wholly solve, particularly related to 51\% attacks, can still be corrected using user-activated soft forks.  Future developments will undoubtedly improve Casper's security and  reduce the need for user-activated soft forks.

\textbf{Future Work.} The current Casper system builds upon a proof of work block proposal mechanism.  We wish to convert the block proposal mechanism to proof of stake.  We wish to prove accountable safety and plausible liveness even when the weights of the validator set change with rewards and penalties.  Another problem for future work is a formal specification of a fork-choice rule taking into account the common attacks on proof of stake.  Future workpapers will explain and analyze the financial incentives within Casper and their consequences.  A particular economic problem related to such automated strategies to block attackers is proving upper bounds on the ratio between the degree of disagreement between different clients and the cost incurred by the attacker.

\textbf{Acknowledgements.} We thank Jon Choi, Karl Floersch, Ozymandias Haynes, and Vlad Zamfir for frequent discussions.

\bibliographystyle{naturemag}
\bibliography{ethereum}

\begin{thebibliography}{10}
\expandafter\ifx\csname url\endcsname\relax
  \def\url#1{\texttt{#1}}\fi
\expandafter\ifx\csname urlprefix\endcsname\relax\def\urlprefix{URL }\fi
\providecommand{\bibinfo}[2]{#2}
\providecommand{\eprint}[2][]{\url{#2}}

\bibitem{shi2017}
\bibinfo{author}{Pass, R.} \& \bibinfo{author}{Shi, E.}
\newblock \bibinfo{title}{Fruitchains: A fair blockchain}.
\newblock In \emph{\bibinfo{booktitle}{Proceedings of the ACM Symposium on
  Principles of Distributed Computing}}, PODC 2017, \bibinfo{pages}{315--324}
  (\bibinfo{publisher}{ACM}, \bibinfo{address}{New York, NY, USA},
  \bibinfo{year}{2017}).
\newblock \urlprefix\url{http://doi.acm.org/10.1145/3087801.3087809}.

\bibitem{dfinity}
\bibinfo{title}{Introducing dfinity crypto techniques} (\bibinfo{year}{2017}).
\newblock
  \urlprefix\url{https://dfinity.org/pdf-viewer/pdfs/viewer.html?file=../library/intro-dfinity-crypto.pdf}.

\bibitem{king2012ppcoin}
\bibinfo{author}{King, S.} \& \bibinfo{author}{Nadal, S.}
\newblock \bibinfo{title}{Ppcoin: Peer-to-peer crypto-currency with
  proof-of-stake} \textbf{\bibinfo{volume}{19}} (\bibinfo{year}{2012}).
\newblock \urlprefix\url{https://decred.org/research/king2012.pdf}.
\newblock
  \bibinfo{note}{\url{https://web.archive.org/save/https://decred.org/research/king2012.pdf}}.

\bibitem{vasin2014blackcoin}
\bibinfo{author}{Vasin, P.}
\newblock \bibinfo{title}{Blackcoin's proof-of-stake protocol v2}
  (\bibinfo{year}{2014}).
\newblock
  \urlprefix\url{http://blackcoin.co/blackcoin-pos-protocol-v2-whitepaper.pdf}.

\bibitem{bentov2016pos}
\bibinfo{author}{Bentov, I.}, \bibinfo{author}{Gabizon, A.} \&
  \bibinfo{author}{Mizrahi, A.}
\newblock \bibinfo{title}{Cryptocurrencies without proof of work}.
\newblock In \bibinfo{editor}{Sion, R.} (ed.)
  \emph{\bibinfo{booktitle}{International Conference on Financial Cryptography
  and Data Security}}, \bibinfo{pages}{142--157}
  (\bibinfo{organization}{Springer}, \bibinfo{year}{2016}).
\newblock \urlprefix\url{http://www.cs.technion.ac.il/~idddo/CoA.pdf}.

\bibitem{castro1999practical}
\bibinfo{author}{Castro, M.}, \bibinfo{author}{Liskov, B.} \&
  \bibinfo{author}{et. al}.
\newblock \bibinfo{title}{Practical byzantine fault tolerance}.
\newblock In \bibinfo{editor}{Leach, P.~J.} \& \bibinfo{editor}{Seltzer, M.}
  (eds.) \emph{\bibinfo{booktitle}{Proceedings of the Third Symposium on
  Operating Systems Design and Implementation}}, vol.~\bibinfo{volume}{99},
  \bibinfo{pages}{173--186} (\bibinfo{year}{1999}).
\newblock \urlprefix\url{http://pmg.csail.mit.edu/papers/osdi99.pdf}.

\bibitem{kwon2014tendermint}
\bibinfo{author}{Kwon, J.}
\newblock \bibinfo{title}{Tendermint: Consensus without mining}
  (\bibinfo{year}{2014}).
\newblock \urlprefix\url{https://tendermint.com/static/docs/tendermint.pdf}.

\bibitem{algorand}
\bibinfo{author}{Chen, J.} \& \bibinfo{author}{Micali, S.}
\newblock \bibinfo{title}{{ALGORAND:} the efficient and democratic ledger}.
\newblock \emph{\bibinfo{journal}{CoRR}}
  \textbf{\bibinfo{volume}{abs/1607.01341}} (\bibinfo{year}{2016}).
\newblock \urlprefix\url{http://arxiv.org/abs/1607.01341}.

\bibitem{nakamoto}
\bibinfo{author}{Nakamoto, S.}
\newblock \bibinfo{title}{Bitcoin: A peer-to-peer electronic cash system}
  (\bibinfo{year}{2008}).
\newblock \urlprefix\url{https://bitcoin.org/bitcoin.pdf}.

\bibitem{minslashing}
\bibinfo{author}{Buterin, V.}
\newblock \bibinfo{title}{Minimal slashing conditions} (\bibinfo{year}{2017}).
\newblock
  \urlprefix\url{https://medium.com/@VitalikButerin/minimal-slashing-conditions-20f0b500fc6c}.

\bibitem{sompolinsky2013accelerating}
\bibinfo{author}{Sompolinsky, Y.} \& \bibinfo{author}{Zohar, A.}
\newblock \bibinfo{title}{Accelerating bitcoin's transaction processing. fast
  money grows on trees, not chains.} \textbf{\bibinfo{volume}{2013}}
  (\bibinfo{year}{2013}).
\newblock \urlprefix\url{http://eprint.iacr.org/2013/881}.

\end{thebibliography}
%\section{References}
%\bibliographystyle{plain}
%\bibliography{ethereum}
%\thebibliography

%\input{appendix.tex}
%\clearpage

%%%%%%%%%%%%%%%%%%%%%%%%%%%%%%%%%%%%%%%%%%%%%%%%%%%%%%

\end{document}